# Challenges and Best Practices in Corporate AI Governance: Lessons from the Biopharmaceutical Industry

**Jakob Mökander,[1,2] Margi Sheth,[3] Mimmi Gersbro-Sundler,[3] Peder Blomgren,[3] Luciano Floridi[1,4]**

[1] Oxford Internet Institute, University of Oxford, 1 St Giles', Oxford, OX1 3JS, UK
[2] Center for Information Technology Policy, Princeton University, Princeton, NJ 08544, US
[3] R&D Data Office, Data Sciences and AI, BioPharmaceuticals R&D, AstraZeneca, Cambridge, UK
[4] Department of Legal Studies, University of Bologna, Via Zamboni 33, 40126, Bologna, Italy

**Correspondence:** < jakob.mokander@oii.ox.ac.uk >

**Keywords:** artificial intelligence, astrazeneca, case study, ethics, governance, implementation, practice

**Abstract**

While the use of artificial intelligence (AI) systems promises to bring significant economic and social benefits, it is also coupled with ethical, legal, and technical challenges. Business leaders thus face the question of how to best reap the benefits of automation whilst managing the associated risks. As a first step, many companies have committed themselves to various sets of ethics principles aimed at guiding the design and use of AI systems. So far so good. But how can well-intentioned ethical principles be translated into effective practice? And what challenges await companies that attempt to operationalize AI governance? In this article, we address these questions by drawing on our first-hand experience of shaping and driving the roll-out of AI governance within AstraZeneca, a biopharmaceutical company. The examples we discuss highlight challenges that any organization attempting to operationalize AI governance will have to face. These include questions concerning how to define the material scope of AI governance, how to harmonize standards across decentralized organizations, and how to measure the impact of specific AI governance initiatives. By showcasing how AstraZeneca managed these operational questions, we hope to provide project managers, CIOs, AI practitioners, and data privacy officers responsible for designing and implementing AI governance frameworks within other organizations with generalizable best practices. In essence, companies seeking to operationalize AI governance are encouraged to build on existing policies and governance structures, use pragmatic and action-oriented terminology, focus on risk management in development and procurement, and empower employees through continuous education and change management.

## 1       Introduction | The need for corporate AI governance

There are two main reasons why artificial intelligence (AI) might be giving business leaders around the world sleepless nights. The first is sheer excitement at the possibilities of the technology.[1] The second is abject fear about the implications of getting it wrong.[2] The potential of autonomous and self-learning technologies to revolutionize industries is only just starting to be realized, with far-reaching implications for human development, economic prosperity, and the financial prospects of individual firms (Floridi et al., 2018). Unsurprisingly, companies are scrambling to implement AI-powered solutions, conscious no doubt that their competitors will be doing the same. But in the rush to join the revolution, there is great danger of organizations making missteps that could lead them into legal and ethical minefields, where they might not only suffer potentially fatal reputational damage but also cause real-world harm (Silverman, 2021).

Consider healthcare as an example of both the benefits and the risks associated with AI. Across the industry, AI systems are already saving lives and increasing life quality by aiding medical diagnostics, driving service improvements through better forecasting, and enabling more effective drug discovery processes (Schneider, 2019; Topol, 2019). However, along with excitement and opportunity, the use of AI systems in healthcare is coupled with serious ethical challenges. Such systems may leave users vulnerable to discrimination and privacy violations (Laurie et al., 2014). They can also erode human self-determination and enable wrongdoing (Tsamados et al., 2021).

Recently, public outcry against specific AI use cases (Holweg et al., 2022) and proposals for "hard" legislation in both the EU[3] and the U.S.[4] have stressed the urgency of addressing these challenges. In that context, defining and communicating ethical principles is a crucial first step toward implementing effective AI governance. It is also a step that many organisations in both the private and the public sector have already taken.[5] However, organisations still lack effective ways of translating those abstract principles into concrete actions that will enable them to achieve the benefits of AI in ways that are ethical, legal, and safe (Gianni et al., 2022; Morley et al., 2019).

---

[1] This bullish mood was reflected in a much-cited report by PwC (2017), which suggested that organisations will be 40% more efficient by 2035 thanks to their adoption of AI systems. While these numbers are debatable, the use of AI systems can lower costs, increase consistency, and enable novel solutions to complex problems (Taddeo & Floridi, 2018).

[2] A survey by McKinsey & Company (2021) found that business leaders worry that their companies lack the capacity to address the full range of risks posed by AI. The same survey also found that executives are particularly concerned about risks relating to cybersecurity, data privacy, and regulatory compliance.

[3] See e.g. The Artificial Intelligence Act (European Commission, 2021).

[4] See e.g. The Algorithmic Accountability Act of 2022 (Office of U.S. Senator Ron Wyden, 2022).

[5] Institutions like the European Commission (AI HLEG, 2019), the OECD (2019) and the IEEE (2019) have all published principles with the aim of guiding the design and use AI systems. In parallel, many companies have chosen to develop and publish their own sets of AI ethics principles.

As a team consisting of both industry practitioners[6] and academic researchers,[7] we set out to fill that gap. Over 12 months, we helped guide and coordinate the roll-out of AI governance within AstraZeneca, a multinational biopharmaceutical company. Throughout the process, we documented the challenges the organisation faced and the best practices that were developed to manage the different tensions that arose. Our findings – which are summarised in this article – include insights that are broadly applicable across industries, offering answers to some of the questions that may be troubling the sleepless manager: What challenges await when implementing AI governance? How can well-intentioned ethical principles be translated into effective practice? And what initial investments are required to reap long-term benefits?

Before proceeding, is should be noted that the pharmaceutical industry – from which our case study is drawn – is well-positioned to pioneer the operationalization of AI governance for three reasons. First, the pharmaceutical industry has a long history of dealing with sensitive data. As a result, governance structures already exist to identify and mitigate technology-related risks. Second, pharmaceutical companies have always operated in an environment governed by laws as well as trust. Indeed, many approaches to AI ethics are based on the classical principles of bioethics (Blasimme & Vayena, 2021). Finally, developing new drugs is not only a data-driven but also a resource-intense endeavour. Globally, over $200 bn is spent on pharmaceutical R&D each year.[8] Hence, there are strong incentives to use AI systems in ways that avoid regulatory red tape.

## 2    Case study | AstraZeneca's AI governance journey

As an R&D-driven organization, AstraZeneca's core business is to use science and innovation to improve health outcomes through more effective treatment and the prevention of complex diseases. In doing so, the company uses AI systems in many different ways. These include biological insight knowledge graphs to improve drug discovery processes, machine-learning powered image recognition software for faster and more accurate medical analysis, and natural language processing models to prioritize adverse event reports (Crowe, 2020; Lea et al., 2021).

All these use cases change how the company collects, analyzes, and utilizes data. And as technologies and ways of working evolve, so must organizational governance. In November 2020, AstraZeneca's board moved toward addressing that need by publishing a set of Principles for Ethical

---

[6] As members of AstraZeneca's R&D Data Office, MS, MSG, and PB have been instrumental in shaping, coordinating, and driving the operationalization of AI governance within the organization.

[7] LF is a professor of philosophy and ethics of information and a former member of the European Commission's High-Level Expert Group on AI. JM is a PhD candidate at the Oxford Internet Institute. Over a period of two years, from Q3 2020 to Q3 2022, LF and JM observed and documented AstraZeneca's internal activities related to AI governance.

[8] www.iqvia.com/insights/the-iqvia-institute/reports/global-trends-in-r-and-d-2022.

Data and AI.[9] Those principles stipulate that the use of AI systems should be private and secure, explainable and transparent, fair, accountable, human-centric, and socially beneficial.

AstraZeneca's ethics principles aim to help employees and partners navigate the risks associated with AI systems. Yet principles alone cannot ensure that such systems are designed and used ethically (Mittelstadt, 2019), and their implementation is never straightforward (Ryan et al., 2021). Moreover, like many other multinational corporations, AstraZeneca is a decentralized organization. Different business areas were thus allowed to develop their own AI governance structures to reflect variations in objectives, digital maturity, and ways of working.

To support and unify local activities, the company launched four enterprise-wide initiatives:
- The creation of overarching *compliance-* and *guidance documents*. The aim thereby was to break down each high-level principle into more tangible and actionable formulations.[10]
- The development of a *Responsible AI playbook*. The aim thereby was to provide detailed, end-to-end guidance on developing, testing, and deploying AI systems within AstraZeneca.[11]
- The establishment of an internal *Responsible AI Consultancy Service,* and an *AI resolution Board*. These new organizational functions were established to (i) facilitate the sharing of best practices, (ii) educate staff, and (iii) monitor the governance of AI projects.
- The commissioning of an *AI audit* conducted in collaboration with an independent party. By subjecting itself to external review, AstraZeneca got valuable feedback on how to improve its existing and emerging AI governance structures.[12]

The above-listed initiatives may appear straight forward. But they only emerged out extended – and sometimes ad hoc – internal processes that came up against both conceptual difficulties and organizational tensions. In the remainder of this article, we highlight the challenges AstraZeneca faced in its efforts to operationalize AI governance and discuss lessons learned, i.e., how these challenges can be managed in pragmatic and constructive ways.

## 3    Practical implementation challenges | What to be prepared for?

Organizations seeking to operationalize AI governance face both conceptual and practical difficulties. Our research and first-hand experience suggest that there are four main challenges:

---

[9] www.astrazeneca.com/sustainability/ethics-and-transparency/data-and-ai-ethics.html.

[10] For example, the principle of transparency was interpreted procedurally, meaning that all business areas must be open about their use of AI systems as well as about the strengths and limitations these systems may have.

[11] The Responsible AI Playbook takes the form of an online repository, which is being continuously updated to direct AstraZeneca employees to relevant resources, guidelines, and best practices.

[12] See Mökander & Floridi (2022) for a descriptive account of how the AI audit was conducted.

## 3.1 Balancing interests

The first challenge concerns the tension between risk management and innovation. The use of AI systems in the pharmaceutical industry gives a stark illustration of that issue. Obviously, the industry must put patients' safety first. Often, that means using available technologies to develop new drugs or to diagnose and intervene early in the course of a disease. To ensure that such drugs are safe, AstraZeneca trains AI systems to detect treatment response patterns (Nadler et al., 2021). But red tape related to AI governance could restrict the development of such potentially lifesaving procedures. In such circumstances, how does a company "err on the safe side?" There is no simple, one-size-fits-all answer. Instead, organizations seeking to operationalize AI governance should prioritize defining and controlling the risk appetite of different projects.

## 3.2 Defining 'AI'

Second, every policy needs to define its scope, but how can you do this when there is no universally accepted definition of AI?[13] Determining the scope of AI governance is especially difficult because the technology is always embedded in larger sociotechnical systems, in which processes driven by humans and by machines overlap (Lauer, 2020). That is why establishing the scope of AI governance is a balancing act. Make the scope overinclusive, and you create unnecessary administrative burdens. Make it underinclusive, and risks will go under the radar (Danks & London, 2017). In AstraZeneca's case, countless meetings were spent on discussing how to best strike that balance. They key to move beyond such discussion is to realize that there is a three-way trade-off between how precisely you define the scope of your AI governance, how easy it is to apply it, and how generalizable it is.

## 3.3 Harmonizing standards

Third, the same requirements must apply to all AI systems used by an organization. If not, corporate AI governance may simply persuade managers to outsource unethical or risky projects (Floridi, 2019). But the drive to impose uniform requirements creates new tensions. Large organizations often comprise distinct business areas that operate independently. The cycle of designing and training AI systems often involves multiple organizations. For example, AstraZeneca collaborates with BenevolentAI, a British start-up, to identify treatments against chronic kidney disease by using the former's rich datasets to build biological insight knowledge graphs.[14] This is not an exception but the rule: AI systems result from supply chains spanning multiple actors and geographic regions (Crawford, 2021). Harmonizing standards means treating all AI systems equally, regardless of whether they have been developed in-house or procured from third parties.

---

[13] See Wang (2019) for an excellent overview of different, partly conflicting, definitions of AI.

[14] www.benevolent.com/news/astrazeneca-starts-artificial-intelligence-collaboration-to-accelerate-drug-discovery.

## 3.4   Measuring results

Fourth, ethics is hard to quantify, and it is not clear how organizations seeking to operationalize AI governance can measure and demonstrate their success. One option is assessing how AI systems operate in terms of fairness, transparency, and accountability. However, it is hard to find ways to quantify and measure these in practice (Kleinberg, 2018). And, as Goodhart's Law reminds us, when a measure becomes a target, it ceases to be a good metric (Strathern, 1997). Alternatively, organizations could focus on designing process-based KPIs to capture the mechanisms in place to mitigate technology-related risks. Yet such checklists tend to reduce AI governance to a box-ticking exercise. Perhaps the solution here comes down to a question of mindset: In an AI governance context, the main purpose of KPIs is not to assess whether a specific system is "ethical" but rather to spark debates about ethics that inform design choices.

## 4   Discussion | Best practices and lessons learned

While the challenges discussed above are real and important, they are not insurmountable. Our experiences from coordinating and observing AstraZeneca's efforts to operationalize AI governance can be condensed into four transferable best practices:

## 4.1   Build on existing policies and governance structures

AI governance is most likely to be effective when integrated into existing governance structures and business processes (Hodges, 2015). Policies that duplicate existing structures may be perceived as unnecessary by the people expected to implement them. Rather than adding steps to their software development processes, organizations can simply update them so that solution requirements align with the objectives of AI governance. This makes is easier for employees to understand what is expected of them and how any new measure related to AI governance impacts their daily tasks. For example, AstraZeneca's software developers and clinical experts found that trying to implement the company's AI ethics principles pushed them to think about their projects in new ways that can help organizations improve their processes and workflows. In short, AI governance is most effectively operationalized when such advantages are clearly communicated.

## 4.2   Use pragmatic and action-oriented terminology

It is less important to define what AI is in abstract terms and more important to establish processes for identifying those systems that require additional layers of governance.[15] Rather than struggling to pin down a precise definition of AI, AstraZeneca created a guidance document that describes the characteristics of the systems to which their ethics principles apply. A list of examples does not constitute a definition of AI, but it can nevertheless help employees determine whether a specific use case is in scope. Also, following the European Commission (2021), AstraZeneca adopted a risk-based approach, classifying systems as either low-, medium- or high-risk with proportionate governance requirements attached to each level. Using a familiar concept as "risk assessment" helps organizations integrate AI governance into their existing quality management processes. A risk-based approach is also future-proof because it avoids the trap of committing to a definition that could become obsolete as the technology rapidly develops.

---

[15] Several recent publications have proposed pragmatic ways of classifying AI systems for the purpose of implementing corporate AI governance. See e.g. Aiken, (2021), Mökander et al., (2022), and OECD (2022).

### 4.3   Focus on risk management in development and procurement

Distinguishing between compliance assurance and risk assurance helps to harmonize standards across organizations. While compliance assurance compares organizational procedures to existing laws and regulations, risk assurance asks open-ended questions about how different business areas work to identify and manage risk. Because regulations vary across jurisdictions and sectors (Viljanen & Parviainen, 2022), it is not always practically feasible to audit all parts of a large, multinational organization for compliance against the same substantive standards. In contrast, risk assurance can be adapted locally to reflect how different business areas understand risk. Because they leave space for managers in different regions and business areas to justify their governance design choices, it is both possible and desirable to subject all parts of an organisation to harmonized AI risk audits. Again, this does not necessarily require the creation of additional layers of governance. Organizations should simply focus on finding any gaps in their existing development and procurement processes and filling them by adding ethics-based evaluation criteria for AI systems.

### 4.4   Empower employees through continuous education and change management

Because corporate AI governance is about change management, internal communication and training efforts are key. In AstraZeneca's case, these efforts were continuous and happened simultaneously on several different levels. For example, the ethics principles were agreed upon through a bottom-up process that included extensive consultations with employees. An important aspect of this process was anchoring the ethics principles with internal stakeholders. After all, ensuring that AI systems are designed and used legally, ethically, and safely requires organizations not only to have the right tools in place but also to make their employees aware of them and willing to use them.

Admittedly, change management is no easy task and the implementation of AI governance is no exemption: Humans have limited attention spans, and employees are frequently bombarded with information about different governance initiatives (Baldwin & Cave, 1999). That said, our first-hand experiences suggest that much can be done to facilitate a successful implementation of AI governance. To start with, communication concerning AI governance is most effective when supported by senior executives. AI governance is also more likely to be implemented when aligned with incentives for individuals and business areas. Put differently, employees must be enabled and supported to do the right thing. That includes training and education as well as channels through which employees can seek escalate issues without fear of being blamed. Finally, tools such as impact assessments and model testing protocols may be developed by individual teams but should be shared widely to encourage the harmonization of practices and prevent the duplication of efforts.

## 5 Concluding remarks | Upfront investments vs. long-term benefits

Our case study of AstraZeneca shows that the most important step toward good corporate AI governance is to ensure procedural regularity and transparency. To do so, organizations do not need to invent or impose new corporate governance structures. For example, while many useful tools such as model cards (Mitchell et al., 2019) and datasheets (Gebru et al., 2018) and methods, like conformity assessments (Floridi et al., 2022), have already been developed, their use is typically neither coordinated nor enforced. That is why the immediate goal of corporate AI governance should be to interlink existing structures, tools, and methods as well as to encourage and inform ethical deliberation through all stages of the AI lifecycle (Mökander & Axente, 2021).

Efforts to operationalize AI governance incur costs, both financial and administrative. To start with, formulating organizational values bottom-up is a time-consuming activity. In AstraZeneca's case, the process of drafting the ethics principles also included multiple consultations with internal executive leaders on strategy and with academic researchers offering external feedback. Since the publication of its ethics principles in 2020, approximately four full-time staff have been working on implementing AI governance across AstraZeneca. Also, in Q4 2021, the company conducted an "AI audit" in collaboration with an independent third party. Added to the costs of procuring that service, AstraZeneca employees invested around 2,000 person-hours in the audit.

To put those numbers into perspective, consider the costs associated with certification and compliance with hard legislation. According to the European Commission, obtaining certification for an AI system in line with the proposed EU legislation on AI will cost on average EUR 20.000, corresponding to approximately 12% of the development cost (Renda et al., 2021). At the same time, one of the main reasons why technology providers engage with auditors is that it is often cheaper and easier to address system vulnerabilities early in the development process. In addition, good AI governance can help organizations improve several business metrics, including data security, brand management, and talent acquisition (EIU, 2020). This shows that – despite the associated costs – businesses have clear incentives to implement effective corporate AI governance structures.

Our discussion in this article has centered on lessons from the pharmaceutical industry. However, AstraZeneca's situation seems highly representative of the many firms that have recently adopted ethics principles for the design and use of AI systems. That is why the challenges and best practices discussed above should be relevant to any organization seeking to operationalize AI governance. It is vital to remember that such governance will not, and should not, replace the need for the designers, operators, and users of AI systems to continuously reflect on the ethics of their actions. Nevertheless, governance that follows the best practices outlined in this article can help organizations manage the ethical risks posed by AI systems while reaping the economic and social benefits of automation.

## 7 Conflict of Interest

MS, MGS, and PB are employees of AstraZeneca plc. JM is a PhD student at the Oxford Internet Institute. His research is supported by AstraZeneca plc.

## 8 Author Contributions

JM and MS share first authorship of this article. All listed authors have made a substantial, direct, and intellectual contribution to the work, and approved it for publication.

## 9 Acknowledgments

The authors want to thank Olawale Alimi, Mihir Kshirsagar, Karen Rouse, Klaudia Jaźwińska, and David Hagan for helpful comments on earlier versions of this manuscript.